\input harvmac.tex
\vskip 2in
\Title{\vbox{\baselineskip12pt
\hbox to \hsize{\hfill}
\hbox to \hsize{\hfill ROM2F-97/56}}}
{\vbox{\centerline{Brane-like States in Superstring Theory}
\vskip 0.3in
{\vbox{\centerline{and the Dynamics of non-Abelian Gauge Theories}}}}}
\centerline{Dimitri Polyakov\footnote{$^\dagger$}
{polyakov@ihes.fr, polyakov@roma2.infn.it}}
\medskip
\centerline{\it Dipartimento di Fisica}
\centerline{\it Universita di Roma ``Tor Vergata''}
\centerline{\it I.N.F.N. - Sezione di Roma ``Tor Vergata''}
\centerline{\it Via della Ricerca Scientifica,1 I-00133 Roma, ITALIA}
\vskip .5in
\centerline {\bf Abstract}
We propose a string-theoretic ansatz describing the dynamics
of SU(N) Yang-Mills theories
in the limit of large N
in $D=4$. The construction uses in a crucial way 
 brane-like states,  open-string non-perturbative
states described by vertex operators of non-trivial ghost cohomologies.
As we have discussed in  previous papers, these physical vertex operators
do not account  for  particle states in the open-string spectrum,
but rather have to do with the dynamics of  extended objects
such as Dp-branes. In this paper, we show their relevance to
the dynamics of gluons.
We show that this string-theoretic
ansatz enjoys 
worldsheet reparametrization invariance and chirality in  loop space,
 satisfies the loop equations and  the criterium for 
quark confinement. According to our proposal, various gauge theories
are described by  string theories with the same classical
action, but with different measures in the functional integral.
The choice of  measure defines the gauge group, as well as the
effective space-time dimension of the resulting gauge theory.

\Date{December 97}
\vfill\eject
\lref\wil{K.G.Wilson, Phys.Rev.D10:2445-2459,1974}
\lref\ampf{A.M.Polyakov, Nucl.Physics {\bf B486}:23-33(1997)}
\lref\amps{A.M.Polyakov, Nucl.Physics {\bf B120}:429-458(1975)}
\lref\amptr{A.M.Polyakov, Phys.Lett.{\bf 59B}:82-84 (1975)}
\lref\ampfr{A.M.Polyakov, hep-th/9711002}
\lref\gross{D.J.Gross,W.Taylor IV, Nucl.Physics {\bf B400:181-210(1993)}} 
\lref\hw{A.Hanany,E.Witten, Nucl.Physics{\bf B492}152-190}
\lref\witf{E.Witten, hep-th/9706109}
\lref\wits{E.Witten, Nucl.Phys {\bf B500}:3-42 (1997)}
\lref\wittr{E.Witten, hep-th/9710065}
\lref\bars{I.Bars, Phys.Rev {\bf D55}2373-2381}
\lref\costas{}
\lref\klebanov{S.Gubser,A.Hashimoto,I.R.Klebanov,J.Maldacena
Nucl.Phys.{\bf B472}(1996) 231}
\lref\green{M.B.Green, Phys.Lett.{\bf B223}(1989)157}
\lref\me{D.Polyakov,hep-th/9705194, to appear in Phys.Rev.D}
\lref\olive{D.Olive, E.Witten, Phys.Lett.78B:97, 1978}
\lref\bern{J.N.Bernstein, D.A.Leites, Funkts. Analis i ego Pril.,
11 (1977)70-77}
\lref\cederwall{M.Cederwall, A.von Gussich,B.E.W.Nilsson,P.Sundell,
A.Westerberg, hep-th/9611159}
\lref\schwarz{M.Aganagic,J.Park, C.Popescu and J.H.Schwarz,
hep-th/9701166}
\lref\bel{A.Belopolsky, hep-th/9609220}
\lref\witten{E.Witten, hep-th/9610234}
\lref\witte{E.Witten, Nucl.Phys.{\bf B463} (1996), p.383}
\lref\myself{D.Polyakov, LANDAU-96-TMP-4, hep-th/9609092,
to appear in Nucl.Phys.{\bf B484}}
\lref\duff{M.J.Duff, K.S.Stelle, Phys.Lett.{\bf {B253}} (1991)113}
\lref\bars{I.Bars, hep-th/9607112}
\lref\grin{M.Green, Phys.Lett. {\bf B223}(1989)} 
\lref\banks{T.Banks, W.Fischler, S.Shenker, L.Susskind, hep-th/9610043}
\lref\seiberg{T.Banks, N.Seiberg, S.Shenker, RU-96-117, hep-th/9612157}
\lref\dougf{M.R.Douglas, hep-th/9604198, Nucl.Phys.Proc.Suppl.41:66-91,1995}
\lref\dougs{M.R.Douglas, hep-th/9604041}
\lref\dougt{M.R.Douglas, hep-th/9409098}
\lref\wsix{E.Witten, hep-th/9710065}
\lref\seiber{N.Seiberg, hep-th/9705221}
\lref\witcomm{E.Witten, hep-th/9507121}
\lref\strom{A.Strominger, hep-th/9512059}
\lref\massimo{M.Bianchi, hep-th/9711201}
\lref\amprepr{A.M.Polyakov, Phys.Lett.82B: 247-250 (1979)}
\centerline{\bf 1.Introduction}
Quark confinement in  non-Abelian gauge theories
is one of the most intriguing problems in  modern physics,
unsolved up to now. While for  gauge theories with compact $U(1)$ 
gauge group an elegant explanation of the
confining mechanism has been found
 ~\refs{\amps}, it is still a
mystery how  confinement appears in gauge theories with essentially
non-Abelian gauge groups.In his
pioneering work K.Wilson ~\refs{\wil} 
 has  shown that  confinement does occur in  non-Abelian 
theories on the lattice, yet unfortunately his method cannot
be applied to address the problem directly in  the continuum limit. 
 The difficulty  has largely to do  with
our lack of understanding of the dynamics of non-Abelian gauge 
theories in general. The straightforward analysis of these non-linear
theories is extremely difficult. One particularly promising and 
potentially powerful approach to the problem is to relate it to
the dynamics of  extended objects, such as strings and p-branes
(see, for example,~\refs{\ampf, \ampfr, 
\amprepr, \gross, \wits, \witf, \wittr},
for  the original as well as for the modern context of this idea)
As for the string-theoretic approach, there
are two basic difficulties with it. The first , and perhaps the
less significant one, has to do with the fact that we still
do not know how to extract gauge theories
with realistic gauge groups from superstring theory. The second,
and perhaps the more fundamental one
is that the ``internal'' gauge symmetries
of Yang-Mills theories and string theory are significantly different:
in string theory there is a worldsheet reparametrization invariance
under diffeomorphisms, i.e. under coordinate transformations with
a positive Jacobian, while in Yang-Mills theories the reparametrizational
invariance of the contour, which defines the Wilson loop, includes 
the transformations which change the orientation of the contour, i.e.
those with  a negative Jacobian as well.
In other words, the reparametrization
invariance in gauge theories is $extended$ and is not restricted to 
diffeomorphisms  ~\refs{\ampf,\ampfr}. Therefore any string theory
that possibly may describe the dynamics of gauge theories
must be a non-standard one, i.e. it must have this special property
of 
extended reparametrization invariance.
Examples of such string theories without gravity
were proposed  in ~\refs{\ampf};
however the problem to find a string-theoretic ansatz describing
gauge theories with particular realistic gauge groups still remains
unsolved.
In this paper, we  propose and study the string-theoretical
ansatz for  $SU(N)$ gauge theories in $D=4$,  explore the confining
properties in these theories and attempt to explain the interplay
between our approach and the idea of elucidating
the properties of gauge theories from brane dynamics.
The crucial element in our construction
will be to observe  the existence of a peculiar class of
soliton-like physical states in the spectrum of open strings,
described by vertex operators of non-trivial ghost cohomologies
~\refs{\me} and the relevance of some of them to the dynamics
of gluons.
Let us recall the criteria necessary for confinement in gauge 
theories. Let $A_m$, $m=1,...,D$ be the vector potential in some
Yang-Mills theory in D dimensions.Let us define the expectation value of
the Wilson loop:
\eqn\lowen{W(C)=<\Psi(C)>=<e^{{\oint_C}A_m{d}X^m}>}
where $C$ is the contour defining the loop, 
 and $X^m$ is a set of space-time coordinates.
The necessary condition for the confinement is that for 
large enough  contours  $W(C)$ vanishes exponentially with the area 
$A$ of the contour:
\eqn\lowen{W(C)\sim{e^{-\lambda{A}}}}
 In the approach that we are going to develop in this paper, 
we aim to present a string-theoretical ansatz for the
SU(N) Yang-Mills theories in the limit of large N 
in $D=4$ by including the set of non-perturbative
open string states ( which we call brane-like states because
 of their apparent p-brane or Dp-brane nature) in the conventional
superstring theory.This peculiar class of states is described
by vertex operators of essentially non-zero ghost numbers
(which cannot be removed by a picture-changing gauge transformation).
Their S-matrix elements between elementary open string states
vanish, however these operators may 
play an important role
in the non-perturbative dynamics of branes ~\refs{\me}.
Examples of these brane-like vertex operators are given
by:
\eqn\lowen{Z_{mn}(k)=e^{-2\phi}\psi_m\psi_n{e^{ikX}}}
and
\eqn\lowen{Z_{m_1...m_5}(k)=e^{-3\phi}\psi_{m_1}...\psi_{m_5}e^{ikX}}
As  follows from superalgebraic arguments 
the zero-momentum parts of these vertex operators are related
to the topological Page charges of a membrane and a fivebrane
respectively.
However, there is another way to understand their  
origin.
Namely, consider the following  Ramond-Ramond vertex operator:
\eqn\lowen{V_{RR}=e^{-{3\over2}\phi}\Sigma^\alpha(z)
\Gamma^{m_1...m_p}_{\alpha\beta}F_{m_1...m_p}e^{-{1\over2}\bar\phi}
\bar\Sigma^\beta(\bar{z})e^{ikX}}
In the presence of a D-brane this vertex operator is defined on
the disc (or, equivalently, on the half-plane),
therefore its holomorphic and anti-holomorphic parts
are no longer independent of each other, because of the boundary.
For this reason,  when the vertex operator (5) approaches the
"edge" of the D-brane, i.e. the boundary where $z=\bar{z}$,
 internal normal ordering must be performed between
$\Sigma$ and $\bar\Sigma$, as well as inside the ghost part.
As a result of such an internal normal ordering the vertex
operator (3) will appear. This is  somewhat similar to the
situation in the NS sector where 
internal normal ordering between $\partial{X}$ and 
$\bar\partial{X}$ in the vertex operator of the graviton
leads to the non-standard dilaton term 
$\sim\int{d^2{z}}R^{(2)}\phi(X)$ in the $\sigma$-model action.
The vertex operators (3),(4) have to do with the dynamics
of branes, but, as we shall argue in this paper, they are
also relevant to the dynamics of gluons.
The idea is that the zero momentum part of the membrane
vertex operator (3) defines, at the same time, the field strength
$F_{mn}$ in a certain non-Abelian gauge theory with 
some (yet unknown)
gauge group:
\eqn\lowen{F_{mn}(z)\equiv\oint{dw\over{2i\pi}}e^{-2\phi}\psi_m
\psi_n(z+w)}
where the dependence on the holomorphic coordinate $z$ defines
 the dependence on the parameter of the Wilson line.
In order to show that $F_{mn}$ defines a meaningful gauge theory
and to find a gauge group to which it corresponds, we have to carry out
the following steps:

1)Propose a suitable definition of the Wilson loop expectation value
in the string-theoretic approach;

2)Show that our string-theoretic ansatz has  
reparametrization invariance
and satisfies the loop space analogue of the Yang-Mills equations ;

3) Show that it
 satisfies the loop equation for non-Abelian gauge theories ;

4)Find the gauge group analyzing the structure of the
loop equation satisfied by proposed superstring ansatz ;

5)Discuss the relevance to $D=4$  in view of the fact that our original
superstring theory is ten-dimensional.

Let us now  begin to implement the above program.
First of all, the Wilson line $\Psi(C)$ is  determined by the 
expression:
\eqn\grav{\eqalign{\Psi(C)=e^{g\oint_C{A_m}{dX^m}}\approx
e^{g\int_C{dz}{d\bar{z}}F_{mn}d\sigma^{mn}}=
e^{g\int{dzd\bar{z}}\theta(C)\partial{X^m}\bar\partial{X^n}(z,\bar{z})
\oint{{dw}\over{2i\pi}}e^{-2\phi}\psi_m\psi_n(z+w)}}}
Here $\theta(C)$ is 
a step function which is equal to 1 inside the contour C and
equal to zero outside,  and $g$ is a coupling constant related
in a certain way  to the bare coupling constant of Yang-Mills
theory; the precise form of this relation will be discussed below. 
Before  giving the definition of the expectation value of the
Wilson line  $W(C)$,  we would like to comment on the reason
why such an average should be computed using the standard 
NSR superstring action (since usually $\Psi(C)$ should be averaged
with respect to the Yang-Mills action $\sim\int{F_{mn}F^{mn}}$).
The justification  is that
 for the $F_{mn}=\oint{dz\over{2i\pi}}e^{-2\phi}\psi_m
\psi_n(z)$ 
of (6) we have the identity:
\eqn\lowen{:\Gamma^4e^{-2\phi}\psi_m\psi_n{e^{-2\phi}}\psi^m\psi^n:(z)
={T_{NSR}}(z)} 
where $T_{NSR}$ is the stress-energy tensor in the NSR superstring 
theory,  $\Gamma^4$ is the normal ordered
fourth power of the picture-changing 
operator :
\eqn\lowen{\Gamma=:e^\phi(G_{matter}+G_{ghost}):=
-{1\over2}e^\phi{\psi^m}{\partial}{X_m}+ghosts}
and $G_{matter}$ and $G_{ghost}$ are the matter and
 ghost worldsheet supercurrents.
The average of the Wilson loop $W(C)$ is then defined as:
\eqn\grav{\eqalign{W(C)=<e^{\int{d^2}z\theta(C)\partial{X^m}
\bar\partial{X^n}(z,\bar{z})\oint{{dw}\over{2i\pi}}e^{-2\phi}
\psi_m\psi_n(z+w)}>_{f(\Gamma)}\cr=
\int{DX}D\psi{D\lbrack{ghosts}\rbrack}f(\Gamma)
exp\lbrace\int{d^2{z}}\partial{X^m}\bar\partial{X^n}(\eta_{mn}\cr+g\theta(C)
\oint{{dw}\over{2i\pi}}e^{-2\phi}\psi_m\psi_n(z+w))+\psi_m\bar\partial
\psi^m+\bar\psi_m\partial\bar\psi^m+S_{ghost}\rbrace}}
Here $f(\Gamma)$ is some function of the picture-changing operator,
regulating the total ghost numbers of the correlation functions
that appear
in the expansion in the gauge coupling constant $g$. This ghost 
number regulator
is needed to make the expression (10)
for the  average $W(C)$ meaningful.Without it, because of
 ghost number conservation, the only non-vanishing correlator
 on the sphere appearing in the expansion
in $g$ would be the one-point function (which is  zero anyway).
Therefore the presence of the ghost number regulator $f(\Gamma)$
in the measure is crucial. Moreover, as we will find out, it is
the precise form of this regulator that determines the gauge group
in the corresponding gauge theory.
Namely,  the proper choice of the measure function $f(\Gamma)$
will ensure that the string-theoretical ansatz (10) for
 the Wilson loop
$W(C)$ satisfies the dynamical equations 
in the loop space  for some gauge theory with
a gauge group G. The  gauge group will be determined from the
structure of  the  loop equations for $W(C)$.
At the same time, a suitable choice for the measure function
$f(\Gamma)$ will enable us to effectively truncate 
the expansion in  the gauge coupling constant $g$, leaving
only a finite number of terms in the series. Therefore the constant
$g$ does not have to be small and the expansion is 
in principle non-perturbative.
Consider a measure function of the type
\eqn\lowen{f(\Gamma)=\sum_n{a(n)}:\Gamma^n:}
where the coefficients $a(n)$ defining the gauge group
 will be determined later.
Expanding  $W(C)$ in $g$ we obtain:
\eqn\grav{\eqalign{W(C){|_{f(\Gamma)}}={\sum_n}{1\over{n!}}g^n
<\theta(C){\prod_{k=1}^n}\int{d^2{z_k}}\oint{{dw_k}\over{2i\pi}}\cr
<\partial{X^{m_k}}\bar\partial{X^{n_k}}(z_k,{\bar{z}}_k):e^{-2\phi}
\psi_{m_k}\psi_{n_k}:(z_k+w_k)>_{f(\Gamma)}}}
where $f(\Gamma)$  implies that
the correlation function is computed with 
the appropriate insertion
 in the functional integration.On the sphere,
due to the ghost number conservation, 
 for every given $n$ in the sum (12) the measure function
$f(\Gamma)$ of (11) may be replaced with 
$f(\Gamma)\rightarrow{a(n-1)}:\Gamma^{4n-2}:$. Then, ignoring $\theta(C)$
in the limit of large $C$ , we find the following the expression
for the average $W(C)$ :
\eqn\grav{\eqalign{W(C){|_{f(\Gamma)}}={\sum_n}
{{a(n-2)g^n}\over{n!}}<{\prod_{k=1}^n}\int{dz_k}
{d{\bar{z}}_k}\oint{{dw_k}\over{2i\pi}}\cr
<:\Gamma^{2n-2}:\partial{X^{m_k}}\bar\partial{X^{n_k}}
(z_k,{\bar{z}}_k):e^{-2\phi}\psi_{m_k}\psi_{n_k}:(z_k+w_k)>}}
Using  the independence of the correlators on the locations
of the picture-changing operators as well as the fact that
${lim_{z\rightarrow{w}}}:\Gamma^2(z)e^{-2\phi}\psi_m\psi_n:(w) 
\sim\psi_m\psi_n(w)+ghosts$ the computation of $W(C)$ in the limit
of large contours gives
\eqn\grav{\eqalign{W(C)|_{f(\Gamma)}=
\sum_n{{a(2n-2)g^n}\over{n!}}{\prod_{k=1}^n}
\int{dz_k}{d{\bar{z_k}}}\oint{{dw_k}\over{2i\pi}}\cr\lbrace
{\prod_{<i_1,j_1=1>}^n}{{{\eta^{m_{i_1}m_{j_1}}
\eta^{n_{i_1}n_{j_1}}(\eta_{m_{i_1}m_{j_1}}\eta_{n_{i_1}n_{j_1}}
+\eta_{m_{i_1}n_{j_1}}\eta_{m_{j_1}n_{i_1}})}\over{{{(z_{i_1}-z_{j_1}) 
}^2}{{({\bar{z}}_{i_1}-{\bar{z}}_{j_1})}^2}{{({\tilde{w}_{i_1}}
-{\tilde{w}_{j_1}})}^2}}}}+permutations\rbrace   }}
where $\eta_{mn}$ is the Minkowski metric and
 ${{\tilde{w}}_{i_1}}\equiv{z_{i_1}}+w_{i_1}$
To compute the expression (14) let us first evaluate the integral
\eqn\grav{\eqalign{I=\int{dz_i}{d\bar{z}_i}\int{dz_j}{d\bar{z}_j}
\oint{{dw_i}\over{2i\pi}}\oint{{dw_j}\over{2i\pi}}
{1\over{{{(z_i-z_j)}^2}{{(\bar{z}_i-\bar{z}_j)}^2}{{(\tilde{w}_i-
\tilde{w}_j)}^2}}}}}
Changing variables according to:
\eqn\grav{\eqalign{u_{1}=z_i-z_j+w_i\cr
u_{2}=z_i+z_j\cr
u_{3}=w_i\cr
u_{4}=w_j}}
for a given $i$
and evaluating the contour integrals in (15) we obtain the following
result
\eqn\lowen{I{\sim}2A({\int_{-\infty}^{\infty}}du{{\delta(u)}\over{u}}-
\lbrack{\int_{-\infty}^{\infty}}du{{(\delta(u))}^2}\rbrack^2)}
where $A$ is the area of the contour $C$.
The divergent integrals in (17) must be regularized.
The regularization scheme is the standard one, based on the
Fourrier representation for the delta-function:
\eqn\lowen{\delta(u)={1\over{2\pi}}{\int_{-\infty}^{\infty}}
{dp}e^{ipu}}
and the identity
\eqn\lowen{{\int_0^{\infty}}du{e^{ipu}}=
lim_{\epsilon\rightarrow{0}}\int_0^{\infty}du{e^{(ip-\epsilon)u}}=
{i\over{p}}}
Then
\eqn\grav{\eqalign{\int_{-\infty}^{\infty}{du}{(\delta(u))}^2=
{2\over{4\pi^2}}\int_{0}^{\infty}du\int{dp_1}{dp_2}
e^{i(p_1+p_2)u}\cr={i\over{2\pi^2}}\int{dp_1}{dp_2}{1\over{p_1+p_2}}
\sim{i\over{\pi^2}}\Lambda({ln}\Lambda-ln{\alpha})}}
where $\Lambda$ is the ultraviolet momentum cutoff and
$\alpha$ is the momentum corresponding to the inverse size of the
contour $C$. For  contours C such that 
$|ln{\alpha}|<<|ln\Lambda|$ that is, for  contours small enough
compared to the absolute value of the cutoff, but still large
enough to ignore the boundary effects in the correlators 
(12),(13), the term with ${ln}\alpha$ may be ignored in the last
formula. Analogously, one may show that
\eqn\lowen{{\int_{-\infty}^{\infty}}du{{\delta(u)}\over{u}}\sim
{1\over\pi}\Lambda({ln}\Lambda-1)}
We see that for $\Lambda\rightarrow\infty$ the second divergent
term in the expression (17) dominates, and
therefore the result of the regularization is given by:
\eqn\lowen{I\sim{2\over{\pi^4}}A{({\Lambda{ln\Lambda}})^2}}
Substituting (22) into the expansion (14) for $W(C)$ and taking
the permutations into account we obtain:
\eqn\grav{\eqalign{W(C)|_{f(\Gamma)}\sim\sum_n{{a(n-2)(n-1)!g^n}\over{n!}}
\lbrack{Ah(\Lambda)}\rbrack^n}}
where $h(\Lambda)\sim(\Lambda{ln}(\Lambda))^2$ is the cutoff
function resulting from the regularization (22).
It is not difficult now to choose the coefficients $a(n)$ in (23)
so that $W(C)|_{f(\Gamma)}$ will have the proper behaviour (2)
of the confining phase of large $N$ Yang-Mills. 
The crucial restriction, however, is
that these coefficients must be chosen so that $W(C)$ satisfies
the loop equations, i.e. the dynamical equations of Yang-Mills theory with
some gauge group, written in the space of loops. 
If this requirement is satisfied, $W(C)|_f(\Gamma)$
with the given choice of $a(n)$ can indeed be identified with the
desired  gauge variable,
i.e. it will define the average of the Wilson 
loop corresponding to some non-Abelian theory. 
The resulting gauge group will be determined from the 
structure of loop equations.
A priori, it is far from obvious that such a choice is possible.
Nevertheless, we shall show that it does really
exist.Namely, we shall argue that the following choice of the measure
function $f(\Gamma)$:
\eqn\lowen{f(\Gamma)\equiv{f(\Gamma,\Lambda)}=
{{(1+\Gamma^2)}\over{N(ln(\Lambda))^2}}e^{-{{\Gamma^4}\over
{N(ln(\Lambda))^2}}}\equiv{Ne_R^4}(1+\Gamma^2)e^{-N(e_R\Gamma)^4}}
in the limit of large $N$ 
connects the $D=10$ string theory to a description of the dynamics
of $SU(N)$ Yang-Mills theory in $D=4$. In the last equation
$e_R$ stands for the renormalized 
Yang-Mills coupling constant in the  $SU(N)$ theory.
The effective change of  space-time dimension will be attributed
to the insertion of $f(\Gamma)$ in the measure of integration,
that corresponds to some configuration of branes (or brane-like states).
 Let us now return to the choice (24) of the measure function $f(\Gamma)$.
 First of all, it is not difficult to show that with this choice
the average $W(C)$ does have the behaviour (2) corresponding to the
confining phase. 
The substitution of (24) into (23) gives:
\eqn\lowen{W(C)|_{f(\Gamma)={{(1+\Gamma^2)}
\over{N(ln(\Lambda))^2}}e^{-{{\Gamma^4}
\over{N(ln(\Lambda))^2}}}}=\sum_n{{(-1)^n}\over{n!}}{\lbrack
{{gh(\Lambda)A}\over{N(ln(\Lambda))^2}}\rbrack}^n=
e^{-{{g\Lambda^2{A}}\over{N}}}}
Note that the ultraviolet cutoff dependence of this result is
 the one predicted by the lattice calculations for  SU(N)
gauge theories in $D=4$ ~\refs{\wil}.
Now we have to show that $W(C)$ with the choice (24) of measure
function is indeed the dynamical variable 
of  the large N limit of a SU(N) gauge theory,
i.e. that  it satisfies the loop equations
corresponding to the SU(N) gauge group:
\eqn\lowen{{{\partial^2{W(C)}}\over{{\partial{X^2(z,\bar{z})}}}}=-e^2
\oint{dY^m}\delta(X(z)-Y)\lbrack{W(C_1,C_2)}-{1\over{N}}W(C)\rbrack
\bar\partial{X_m}(z,\bar{z})}
Here $C$ is the contour with one self-intersection; the 
self-intersection point divides into two contours $C_1$ and $C_2$
and $W(C_1,C_2)\equiv<\Psi(C_1)\Psi(C_2)>$ and the second derivative
with respect to $X$ is defined as follows:
\eqn\lowen{{{\partial^2{W(C)}}\over{{\partial{X^2(z,\bar{z})}}}}\equiv
{lim}_{\epsilon\rightarrow{0}}\int_{|\alpha|<\epsilon}{d^2}\alpha
{{\delta{W(C)}}\over{X^m(z)}}{{\delta{W(C)}}\over{X_m(z+\alpha)}}}
Before doing that, however, let us show first that
the $classical$ variable $\Psi(C)$ of (7) satisfies
the following relations
\eqn\lowen{\partial{X^m}(z){{\delta\Psi(C)}\over{\delta{X^m}(z)}}=0,}
equivalent to the reparametrization invariance, together with
\eqn\lowen{{{\delta^{2}\Psi(C)}\over{\delta{X^m}(z)\delta{X^n}({w})}}
-{{\delta^2\Psi({{C}})}\over{\delta{X^n}({z})\delta{X^m}(w)}}+
\lbrack{{\delta\Psi(C)}\over{\delta{X^m}(z)}},
{{\delta\Psi(C)}\over{\delta{X^n}(w)}}\rbrack=0,}
 chirality in the loop space; and finally
\eqn\lowen{{{\delta}\over{\delta{X^m(z)}}}({{\delta\Psi(C)}\over{\delta
{X_m}(z)}}\Psi^{-1}(C))=0,}
 Yang-Mills equations written in the loop space formalism.
Let us now prove that 
the superstring ansatz (7) for $\Psi(C)$ satisfies all these 
identities.

For convenience, let us denote $L_m(C,z)\equiv{{\delta\Psi(C)}\over
{{\delta}X^m(z)}}$.
On the other hand, one may show that on-shell, due to the NSR equations
of motion, one has
\eqn\lowen{{\bar{L}}_m(C,\bar{z})=0.}
The proof of (28) follows from
\eqn\lowen{\bar\partial{X^m}L_m=g\lbrace{lim}_{u\rightarrow{z}}
{{\eta^{mn}}\over
{(\bar{z}-\bar{u})^2}}+:\bar\partial{X^m}\bar\partial{X^n}:(z,\bar{z})
\rbrace:{F_{mn}}(X(z))\Psi(C):\equiv{0}}
since in the last formula
$F_{mn}$ is multiplied by an expression that is symmetric in the indices
$m$ and $n$.
Next, the condition (29) of chirality in the loop space is fulfilled
due to the fact that
\eqn\grav{\eqalign{{{\delta{L_m}(z,C)}\over{{\delta}X^n(w)}}-
{{\delta{L_n}(z,C)}\over{{\delta}X^m(w)}}=g^2\bar\partial{X_\rho}(\bar{s})
\bar\partial{X_\sigma}(\bar{w}):\lbrack{F_{n\sigma}}(X(w)),F_{m\rho}
(X(z))\rbrack\Psi(C):\cr\equiv{-}\lbrack{L_m},L_{n}\rbrack}}
Finally, the loop space analogue (30) of the Yang-Mills equations 
is satisfied by the ansatz (7), because
\eqn\grav{\eqalign{{{\delta}\over{\delta{X^m}(z)}}
({{\delta\Psi(C)}\over{\delta{X_m}(z)}}\Psi^{-1}(C))=
{{\delta{L^m(z,C)}}\over{\delta{X^m(z)}}}\Psi^{-1}(C)+
L^m(z,C){{\delta\Psi^{-1}(C)}\over{\delta{X^m(z)}}}\cr=-g^2\lbrace
{{lim}_{\bar{u}\rightarrow{\bar{z}}}}
{{\eta^{\rho\sigma}}\over{(\bar{z}-\bar{u})^2}}+
:\bar\partial{X^\rho}\bar\partial{X^\sigma}:(z,\bar{z})\rbrace\times\cr
\times:(\lbrack{F_{m\rho}}(X(z)),F_{m\sigma}(X(u))\rbrack-
\lbrack{F_{m\rho}}(X(z)),F_{m\sigma}(X(u))\rbrack)\Psi(C)\Psi^{-1}(C):
\equiv{0}}}
This ends the proof of (28)-(30).We thus see that 
the expression (7) for $\Psi(C)$  indeed corresponds to
 the dynamical variable in some classical
 gauge theory, i.e. to the Wilson loop
 in some Yang-Mills theory (with a gauge group
 yet to be identified).
Now we have to show that,
for the choice (24) of the measure function $f(\Gamma)$  the 
superstring ansatz  (10)
describes the
gauge theory dynamics in the quantum sense as well; that is, that
the average (10) of $\Psi(C)$ corresponds to the expectation
value of the Wilson loop in
 the $D=4$ $SU(N)$ Yang-Mills theory in the limit of large N.
To do this we first of all have to show that
the expression (10) for $W(C)$ with the choice (24) for the measure
function $f(\Gamma)$ satisfies the loop equation (26) for 
$SU(N)$ gauge theories.
 Let us compute the second derivative of $W(C)$. According to the definition
(27), we have:
\eqn\grav{\eqalign{{{{\partial^2}W(C)}\over{\partial{X^2}(z,\bar{z})}}
|_{f(\Gamma)}
=\cr={{lim}_{\epsilon\rightarrow{0}}}
{\int_{|\alpha|<\epsilon}}d^2{\alpha}<\bar\partial{X_\rho}(\bar{z})
\bar\partial{X_\sigma}(\bar{z}+\bar\alpha)
:F_{n\sigma}F_{m\rho}:(X(z))\Psi(C)>_
{f(\Gamma)}}}
Unfortunately, since the ansatz (6) for the field strength has an
essentially non-zero ghost number, differentiation with respect to $X$
inevitably changes the ``balance of ghosts'' 
in the correlation functions.
In order to compensate this change the differentiation must be accompanied
by the appropriate picture-changing transformation.Thus, the correct
physical variable to consider is 
$:\Gamma^4{{{\partial^2}W(C)}\over{\partial{X^2}}}:|_{f(\Gamma)}$,
rather than simply ${{{\partial^2}W(C)}
\over{\partial{X^2}}}|_{f(\Gamma)}$.
Expanding $\Psi(C)$ in $g$ in a way similar to (13), using the identity:
$\partial{X(z)}\bar\partial{X(\bar{w})}\sim\delta^{(2)}(z-w)$
and  multiplying both sides of the last equation by $:\Gamma^4:$
we obtain:
\eqn\grav{\eqalign{:\Gamma^4{{\partial^2{W(C)}}\over
{\partial{X^2}(z)}}:=\cr={lim}_{\epsilon\rightarrow{0}}
<{\sum_j}\int_{|\alpha|<\epsilon}^{}
d^2\alpha\int{d^2}{u_j}\delta^{(2)}(z+\alpha-u_j)\eta^{\sigma{m_j}}
\bar\partial{X^\rho}(\bar{z})\bar\partial{X^{n_j}}(\bar{u_j})\cr
\times{\Gamma^4}f(\Gamma)
F_{m_jn_j}(X(u_j))F_{m\sigma}(X(z+\alpha))
F_{m\rho}(X(z))\cr\lbrace\sum_n{1\over{n!}}{\prod_{k=1,k\neq{j}}^n}
\int{d{z_k}}d{{\bar{z}}_k}\oint{{dw_k}\over{2i\pi}}
\partial{X^{m_k}}\bar\partial{X^{n_k}}(z_k,\bar{z}_k)
e^{-2\phi}\psi_{m_k}\psi_{n_k}(z_k+w_k)\rbrace>+...
}}
where  we have dropped  terms that vanish as $\epsilon\rightarrow{0}$,
i.e. only the terms with  $\alpha$ in the 
argument of the delta-function are retained.
Evaluation of the integral over $\alpha$ 
then leads to the following expression 
for the second derivative of $W(C)$ with the measure function (24):
\eqn\grav{\eqalign{\Gamma^4{{\partial^2{W(C)}}\over{\partial{X^2}(z)}}=
\cr=g^2<{\sum_j}\int{d}{\bar{u_j}}\delta(X(\bar{u_j}))-X(\bar{z}))
\bar\partial{X_\rho}(\bar{z})\bar\partial{X_{n_j}}(\bar{u_j})
\eta_{n_j\rho}:F_{m\sigma}F_{m\sigma}:(X(z))\cr\times\lbrace\sum_n{1\over{n!}}
\prod_{k=1}^{n-1}\int{d^2}z_k\oint{{dw_k}\over{2i\pi}}:\Gamma^{2}
f(\Gamma):\partial{X^{m_k}}\bar\partial{X^{n_k}}(z_k,\bar{z_k})
e^{-2\phi}\psi_{m_k}\psi_{n_k}(z_k+w_k)\rbrace>=\cr={{2Tr(\eta)g^3}
\over{N(ln\Lambda)^2}}
\int{dY^\rho}\partial{X_\rho}(\bar{z})\delta(Y-X(\bar{z}))
{\sum_{a,b=1}^{n-1,n}}\int{dz_a}{d\bar{z}_a}\int{dz_b}{d\bar{z}_b}\delta^{(2)}
(z_a-z)\delta^{(2)}\cr\times\lbrack(\delta^{(2)}(z_a-z_b))^2
-{{\delta^{(2)}(z_a-z_b)}\over{z_a-z_b}}\rbrack\cr\times
 \sum_n{{g^{n-2}}\over{{\lbrack}N(ln\Lambda)^2\rbrack^{n-2}
n!}}{\prod_{k=1}^{n-2}}<\int{d^2{z_k}}
\oint{{dw_k}\over{2i\pi}}\partial{X_{m_k}}\bar\partial{X_{n_k}}
e^{-2\phi}\psi_{m_k}\psi_{n_k}>\rbrace}}
where $:F_{m\sigma}F_{m\sigma}:(X(z))\equiv:\oint{{dw_1}\over{2i\pi}}
\oint{{dw_2}\over{2i\pi}}e^{-2\phi}\psi_m\psi_\sigma(z+w_1)e^{-2\phi}\psi_m
\psi_\sigma(z+w_2):$ and we have used the O.P.E. rule
$:\Gamma^2F_{m_1n_1}:(u)F_{m_2n_2}(z)\sim
\delta^{(2)}(u-z)\eta_{m_1m_2}F_{n_1n_2}(z)+...$
and the expression for the two-point correlation function:
\eqn\grav{\eqalign{<\Gamma^2\partial{X^a}\bar\partial{X^b}F_{ab}(z,\bar{z})
\partial{X^c}\bar\partial{X^d}F_{cd}(u,\bar{u})>=(Tr{\eta})^2
((\delta^{(2)}(z-u))^2-{{\delta^{(2)}(z-u)}\over{z-u}})}}
To see that the loop equation (26) is satisfied we have to show that
the right-hand side of (37) is identical to the right-hand side of
(26).Let us therefore compute $W(C_1,C_2)$,  with the measure
function (24).
We have:
\eqn\grav{\eqalign{W(C_1,C_2)=
<\Gamma^2e^{\theta(C_1)\int{d^2}{z_1}\partial{X^{m_1}}
\bar\partial{X^{n_1}}F_{mn}({z_1},\bar{z})}e^{\theta(C_2)\int{d^2}z
\partial{X^{m_1}}\bar\partial{X^{n_1}}F_{m_2n_2}({z_2},\bar{z_2})}>=\cr=
\sum<{{\theta(C_1)\theta(C_2)g^{n+m}}\over{m!n!}}\Gamma^2f(\Gamma)
{\prod_{k,l=1}^{n,m}}
\int{d^2{z_k}}\partial{X^{m_k}}\bar\partial{X^{n_k}}F_{m_kn_k}(z_k,
\bar{z}_k)\cr\times\int{d^2}{z_l}\partial{X^{m_l}}\bar\partial{X^{n_l}}
F_{m_ln_l}(z_l,\bar{z}_l)>=\cr=
{{{2g}\over{N(ln\Lambda)^2}}}
\sum_m{1\over{n!}}{\prod_{k=1}^{n}}{\sum_{a=1}^n}<\int{d^2}u\int{d^2}{z_k}
\delta^{(2)}(z-u)((\delta^{(2)}(u-z_a))^2-\cr-{{\delta^{(2)}(u-z_a)}\over
{u-z_a}})f(\Gamma){{g^{n+1}}\over{n!}}\partial{X^{m_k}}\bar\partial{X^{n_k}}
F_{m_kn_k}(z_k,\bar{z_k})>+...}}
where we have dropped the terms which become small in the limit
of the large N (with the choice (24) of the measure function
$f(\Gamma)$).We have also used of the fact that in the
terms proportional to $\sim\theta(C_1)\theta(C_2)$ in (39)
the interaction between two given vertices located
at the different halves $C_1$ and $C_2$ of the contour
$C$ occurs only when one of the vertices passes
through the self-intersection point  - this is the origin
of the delta-function    $\delta^{(2)}(z-u)$ in (39).
Comparing the expression (39) for $W(C_1,C_2)$ with the
right-hand side of (37) we see that,
up to a picture-changing transformation 
that restores the correct ghost balance, the second derivative of
$W(C)$ is equal to 
\eqn\lowen{\Gamma^4{{\partial^2{W(C)}}\over{\partial{X^2(z)}}}=
g^2\int{dY^\rho}\bar\partial{X_\rho}
\delta(Y-X(z))W(C_1,C_2)}
The loop equation (26) for the SU(N) Yang-Mills theory
is thus satisfied by the superstring ansatz (10) 
with the measure function (24) in the limit
$N\rightarrow\infty$. The important question, however, is
 why the gauge theory described by the ansatz (10)
is   four-dimensional, while the original string
theory is in $D=10$. Our conjecture is that introducing 
the function $f(\Gamma)$ in the measure of the functional
integration effectively changes the dimension of the space-time
in which the theory lives. First of all, note that the cutoff dependence
of W(C) in (25) is the same as in four dimensions, obtained from
the computation on the lattice ~\refs{\wil}.
The reason for the change of space-time dimensionality may be understood
as follows.The inclusion of $f(\Gamma,\Lambda)$ in the measure is equivalent
to introducing boundaries (or a curvature singularity) on the worldsheet
which, in turn, leads to brane-like effects.
Thus the inclusion of the brane-like states (3), (4) along with some
particular choice of $f(\Gamma,\Lambda)$ is equivalent to a certain
 configuration of branes in space-time.
Some of such brane configurations give rise to the gauge
theories in lower space-time dimensions.
There is another, even more heuristic, way to understand the
conjectured reduction of the space-time dimensionality.
That is, because of the ghost number anomaly cancelation condition
the choice of $f(\Gamma,\Lambda)$ imposes strong restrictions on
the possible worldsheet topologies 
that appear in the expansion in $g_{string}$. 
For instance, the choice (24) of the measure function
 in the limit of large N effectively
suppresses the contributions from all
non-trivial topologies, allowing us to consider the correlation
functions on the sphere only.  As a result, the set of all possible
loop configurations is reduced as well. In turn, as a result of such a
reduction, the loop space including the loop configurations in the
$D=10$ space-time is reduced to that of loop configurations
in the space-time of a lower dimension.
To deduce the effective dimensionality of the space-time for a given 
$f(\Gamma,\Lambda)$
one has to compute the average of the Wilson loop $W(C)_{f(\Gamma)}$
provided that it satisfies the loop equation, otherwise the given choice
of the measure function cannot be related to any physical gauge theory).
Then the cutoff dependence of the result must be compared to the dependence
on the ultraviolet cutoff parameter in
 the corresponding gauge theory in $D$ dimensions,
that follows from the computations on the lattice.
Roughly speaking, given the measure function $f(\Gamma,\Lambda)$,
the dependence on $\Gamma$ determines the gauge group, while
the dependence on $\Lambda$  fixes the effective space-time dimensions of 
the theory.This statement is not quite precise, of course,
since the dependence on $\Gamma$ plays a role in fixing the
effective space-time dimension as well.             
According to the behaviour (25) of the expectation value of the Wilson loop
W(C), the measure function (24) presumably describes the confining phase of
 SU(N) Yang-Mills in four dimensions. It is 
also possible to choose the measure function $f(\Gamma)$ so that
$W(C)$, while still satisfying the large N limit of the loop equation
(26), ,would behave like like $W(C)\sim{e^{-\lambda\Lambda^2{A^{1\over2}}}}$,
corresponding to either the Coulomb or  the magnetic phase
of large N  Yang-Mills in $D=4$. 
Namely with the measure function
\eqn\lowen{f(\Gamma,\Lambda)={1\over{N(ln(\Lambda))^2}}
e^{-{{\Gamma^2}\over{N(ln(\Lambda))^2}}}\equiv{N}{e_R^4}
e^{-N{e_R^4}\Gamma^2}}
a computation similar to the one in (12)-(25) gives
\eqn\lowen{W(C)|_{f(\Gamma)}\sim{e^{-{{g\Lambda^2\sqrt{A}}\over{N}}}},}
 the behaviour which characterizes the Coulomb or the magnetic
phase of Yang-Mills theory.
The element which is still missing in the construction,
and which is necessary to be able to distinguish
between Coulomb and magnetic phases as well as to give a full proof
of the confinement conjecture for the choice
(24) of the measure function, is the expression for the QCD order
parameter,  the ``magnetic moment'' M.
This order parameter is known to satisfy the relation
\eqn\lowen{M\Psi(C)=e^\varphi{\Psi(C)}M}
 where $\varphi$ is the phase factor. 
In the confining phase it should behave as 
\eqn\lowen{M\sim{e^{-\lambda\sqrt{A}}}}
So far we were unable to find an exact 
relation of this parameter to any brane-like
state (3), (4).The natural suggestion is that it should be related
to the magnetic-type fivebrane state (4) according to
\eqn\lowen{M\sim{exp}\lbrace\int{d^2z}\partial
{X_{m_1}}\bar\partial{X_{m_2}}(z,\bar{z})\oint{{dw}\over{2i\pi}}e^{-3\phi}
\psi_{m_3}...\psi_{m_7}{F_{RR}^{m_1...m_7}}(z+w)+permut.\rbrace}
where $F_{RR}$ is the Ramond-Ramond $7$-form field strength;
or the similar expression in the $+1$-picture, with $e^{-3\phi}$
replaced with $e^\phi$.
It appears that the order parameter in the $+1$-picture does satisfy
the property
\eqn\lowen{L^{(+1)}W(C)=W(C)L^{(+1)}}
since the vertices 
$\oint{{dz}\over{2i\pi}}e^\phi\psi_{m_1}...\psi_{m_5}(z)$ and
$\oint{{dw}\over{2i\pi}}e^{-2\phi}\psi_m\psi_n(w)$ commute trivially;
this result is  sensible in the $N\rightarrow\infty$ limit.
It will be important to show that the parameter $M$   
has the appropriate behaviour in the confining phase (with the
measure function (24)) as well as  to study its asymptotic behaviour
for other choices of measure functions that  we hope may
describe the dynamics of Yang-Mills in Coulomb and magnetic phases.
We hope to complete such a computation in the future.

\centerline{\bf Conclusion}

We have argued in this paper that the brane-like states
(3), (4) play a significant role in building a relation
between string theory and non-Abelian gauge dynamics.
The insertion of vertices (3), (4) is equivalent to 
introducing branes; the choice of the measure function 
$f(\Gamma)$ regulates the mutual position of branes.
There is no question that our results are still very far
from being complete and suffer a number of serious drawbacks.
First of all, in the result (25) for the expectation value
of the Wilson loop, while the dependence
on the cutoff is in agreement with 
what one would expect from lattice arguments,
 the dependence on the bare
coupling constant does not reproduce the result  of the
computation on the lattice in $D=4$, that is, in the
formula (25) $g$ appears rather than $ln(g)$. 
Our arguments regarding the
conjectured reduction of the effective 
space-time dimension due to the insertion of $f(\Gamma)$
in the measure
appear to be quite heuristic and intuitive at the present time.
In order to make the ground for these arguments more solid
 one has to present a precise analysis of
the relation between a given choice of measure
function and the ``induced'' brane configuration.
One also has to explore the mechanism of supersymmetry
breaking produced by this insertion (since the resulting gauge
theory is not supersymmetric).
The supersymmetry-breaking term in the action is proportional to
$\sim\int{d^2{z}}{ln}(f(\Gamma(z)))\delta^{(2)}(z-w)$ where
$w$ is  a location of  picture-changing operator.
In spite of the above drawbacks,
we feel that the properties satisfied by 
the  gauge-theoretic observable (7) constructed
 out of brane-like vertices (3), (4), such as the
large N limit of the loop equation (26), as well as the
confining behaviour (25), give  a certain evidence
for the role of the brane-like states in QCD. 
The brane-like states appear as a result of the internal normal
ordering inside the Ramond-Ramond vertex operators,
often ignored in calculations involving  RR states but,
 in our opinion,  crucial when one studies the properties of 
RR vertex operators on surfaces with boundaries (such as D-branes). 
Another possible consequence of this internal normal ordering is
the worldsheet interpretation of
the gauge symmetry enhancement which occurs when 
several  D-branes are joined together ~\refs{\dougf, \dougs}
That is, when several D-branes, each carrying  Ramond-Ramond charge,
come together, one may need to perform the normal ordering inside the
corresponding Ramond-Ramond vertices, as they approach each other.
Another interesting implication for the proposed formalism that relates 
 brane-like states and  measure functions to   gauge groups
would be to try to give a string-theoretic description 
of the recently  constructed six-dimensional string-like theories
without  dynamical gravity
~\refs{\wsix, \seiber, \strom, \witcomm, \massimo}. These theories
arise from either specific configuration of fivebranes in M-theory 
or from  IIB in an ALE background, and 
are known to reduce in the low-energy limit to 
six-dimensional gauge theories (with either U(N) or SU(N) gauge groups)
decoupled from supergravity.
The microscopic description of these theories, however, is still obscure.
It will be interesting to find a class of measure functions that
generate  ALE-type  singularities in the background, and then to show
by explicit computations that gravitons do not in fact arise or become massive
in the spectrum of such  superstring theories
(originally in $D=10$ with $f(\Gamma)$ effectively reducing them to $D=6$.)
In view of the fact that gravitons are in general responsible for
breaking the extended reparametrization symmetry in usual superstring theories
in ten dimensions, one may expect that the recently discovered
string-like theories without gravity in $D=6$ are in fact invariant
under extended reparametrizations and satisfy dynamical
equations similar to large N loop equations 
(in the limit $N\rightarrow\infty$). 
This is another place where the interplay between non-Abelian
gauge theories and brane-like states may appear.
Hopefully, a deeper analysis of these problems may also
bring more  understanding of U-duality in terms of
the worldsheet physics.
\centerline{\bf Acknowledgements}
This research is supported by the European Post-Doctoral Institute
(EPDI). The author acknowledges the kind hospitality of the Institut
des Hautes Etudes Scientifiques (IHES) and the Universita di Roma 
``Tor Vergata'', and especially extends his gratitude to 
T.Damour and A.Sagnotti.
I would also like to thank 
M.Bianchi, F.Fucito, A.M.Polyakov and A.Sagnotti for
many helpful discussions and suggestions.

\listrefs
  
\end